\documentclass[10pt,twocolumn,prl,aps,amsmath,amssymb,floatfix,superscriptaddress,longbibliography]{revtex4-2}

\usepackage{graphicx}% Include figure files
\usepackage{dcolumn}% Align table columns on decimal point
\usepackage{bm}% bold math
\usepackage[utf8]{inputenc}
\usepackage[T1]{fontenc}
\usepackage{mathptmx}
\usepackage{etoolbox}
\usepackage{siunitx}
\usepackage{chemformula}
\usepackage{braket}
\DeclareSIUnit[number-unit-product = {\,}]
\cal{cal}

\usepackage{hyperref}

\begin{document}

%\preprint{AIP/123-QED}

\title{Quantum phases in endofullerene zigzag chains}
% Force line breaks with \\
%\author{Tobias Serwatka}
%\author{Pierre-Nicholas Roy}%
% \email{pnroy@uwaterloo.ca.}
%\affiliation{ 
%Department of Chemistry, University of Waterloo, Waterloo, Ontario N2L 3G1, Canada
%}

\author{Tobias Serwatka}
\affiliation{Department of Chemistry, University of Waterloo, Ontario, N2L 3G1, Canada}
\author{Muhammad Shaeer Moeed}
\affiliation{Department of Physics, University of Waterloo, Waterloo, Ontario, N2L 3G1, Canada}
\affiliation{Institute For Quantum Computing, University of Waterloo, Waterloo, Ontario, N2L 3G1, Canada}
\affiliation{Perimeter Institute for Theoretical Physics, Waterloo, Ontario N2L 2Y5, Canada}
\author{Roger G. Melko}
\affiliation{Department of Physics, University of Waterloo, Waterloo, Ontario, N2L 3G1, Canada}
\affiliation{Perimeter Institute for Theoretical Physics, Waterloo, Ontario N2L 2Y5, Canada}
\author{Pierre-Nicholas Roy}
\affiliation{Department of Chemistry, University of Waterloo, Ontario, N2L 3G1, Canada}
\affiliation{Institute For Quantum Computing, University of Waterloo, Waterloo, Ontario, N2L 3G1, Canada}
\affiliation{Perimeter Institute for Theoretical Physics, Waterloo, Ontario N2L 2Y5, Canada}
\email{pnroy@uwaterloo.ca}

\date{\today}% It is always \today, today,
             %  but any date may be explicitly specified

\begin{abstract}
We employ large-scale density matrix renormalization group calculations to study the quantum phases of dipolar molecules confined in bent (zigzag) endofullerene chains, as a function of the chain angle $\gamma$. For LiF, ferroelectric order persists across the full range $60^\circ < \gamma < 180^\circ$, with the critical effective dipole moment increasing as the chain bends and parallel alignment becomes less favorable. Near the equilateral configuration ($\gamma = 60^\circ$), geometric frustration drives a transition to an antiferroelectric N\'{e}el-ordered phase in which neighboring dipoles anti-align along the chain axis. We show that capturing this reorientation requires including dipolar couplings beyond the nearest-neighbor approximation, since next-nearest-neighbor interactions become equally strong at $\gamma = 60^\circ$. For confined water, o-D$_{2}$O reproduces both ordered phases, whereas p-H$_{2}$O---owing to its large rotational constants---develops no order at any chain angle despite the enhanced coordination of the bent geometry. Because a zigzag chain is the narrowest stripe of a two-dimensional lattice, these results suggest that engineered endofullerene layers could host a rich variety of dipole-ordered quantum phases beyond the ferroelectric ordering observed in previous work.
\end{abstract}

\maketitle

\begin{figure}
\includegraphics[width=\columnwidth]{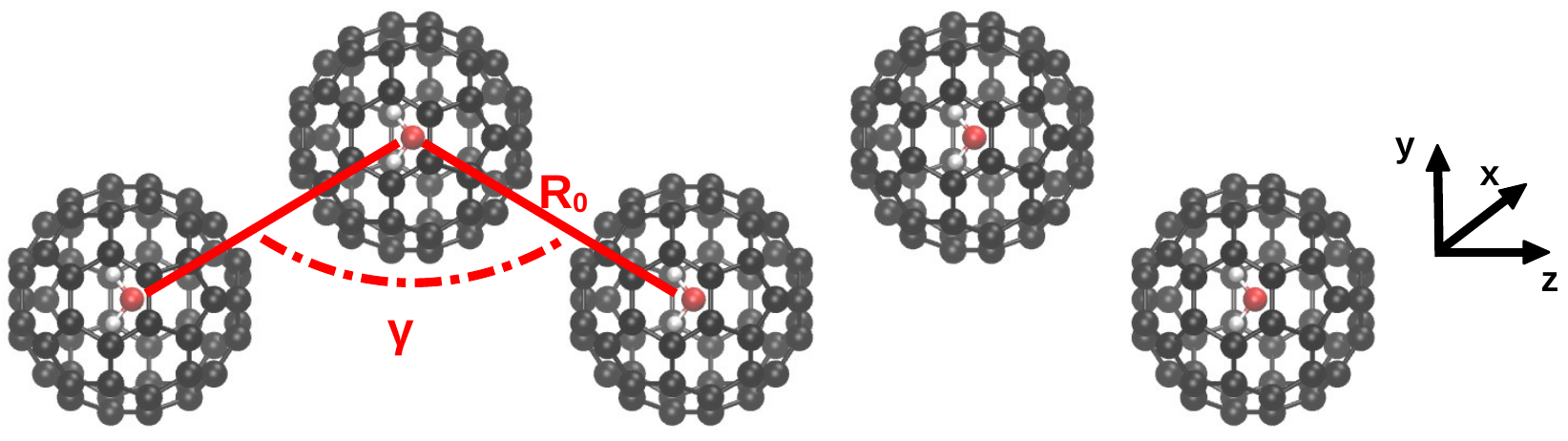}
\caption{Chain of endofullerene units. The configuration of the fullerene chain depends on $R_{0}$, the distance of the centers of mass of neighboring cages and the chain angle $\gamma\in (60^\circ,180^\circ)$. At $180^\circ$ a linear chain is formed and at $60^\circ$ the three nearest-neighbor cages form an equilateral triangle.}
\label{fig:zigzag}
\end{figure}

Nano-confined molecules are a versatile platform for realizing many-body quantum effects \cite{serwatka2023endo, ma2017quasiphase}, from the anomalous transport of nano-confined water \cite{water_transport_prb, hwang_water_self_diffusivity, Hadi_review_nano_transport} to dipole-ordered (ferroelectric) and disordered (paraelectric) phases of polar molecules in endofullerene chains \cite{serwatka2023quantum}. Indeed, dielectric spectroscopy of water confined in beryl and cordierite nanochannels reveals incipient ferroelectricity, quantum paraelectricity, 
and associated phase transitions between these ordered states 
\cite{gorshunov2016incipient,belyanchikov2020dielectric,zhukova2019quantum,belyanchikov2022fingerprints,belyanchikov2022single,gorshunov2022effect,kolesnikov2016quantum}, establishing such assemblies as a tunable platform for studying  quantum criticality  \cite{sachdev2011quantum,vojta2003quantum,bitko1996quantum,coldea2010quantum}.
The cages inhibit hydrogen bonding so the guests interact through their dipoles \cite{serwatka2024planarcriticality}, and ordering arises from spontaneous breaking of the dipole reflection symmetry at zero temperature once the lattice spacing is small enough \cite{serwatka2023quantum,takei2024effective, hodak2003systems}. Beyond linear chains, bent (quasi-1D) carbon nanostructures can also be realized experimentally \cite{chen2015temperature, warner2008rotating, troche2007structural, peng2025monolayer}; Fig.~\ref{fig:zigzag} shows such a configuration with chain angle $\gamma$ and neighbor spacing $R_0$. Dipoles in this geometry are frustrated, as no local orientation satisfies all interactions on a triangular sub-lattice. Such frustration is a central theme in correlated quantum matter, where competing interactions on triangular geometries stabilize unconventional ordered and disordered ground states \cite{abalmasov2018screening,demille2002quantum}.

Here, we study this frustrated lattice with large-scale density matrix renormalization group (DMRG) \cite{white1992density,schollwock2011density,fishman2022itensor} calculations as a function of the chain angle $\gamma$, with diatomic LiF linear rotor and the o-D$_2$O and p-H$_2$O asymmetric top species confined in the cages. Such confined-rotor assemblies have also been treated by exact diagonalization \cite{halverson2018quantifying,HALVERSON201537}, path-integral ground-state Monte Carlo \cite{sahoo2020path,sahoo2021path,zhang2025pimc}, and neural-network states \cite{serwatka2024rnn}, reflecting the challenge of studying such continuous rotational degrees of freedom.

A chain of $N$ rigid endofullerene units can be described by
\begin{align}
\hat{H}=&\sum_{i=1}^{N}\left(-\frac{1}{2M}\Delta_{\vec{R}_{i}}+\hat{T}_{\mathrm{rot},i}+\hat{V}_{\mathrm{guest-cage}}^{i}\right)\notag\\
&+\sum_{[ij]}\hat{V}_{\mathrm{dd}}^{ij}(\Omega_{i},\vec{R}_{i},\Omega_{j},\vec{R}_{j};R_{0},\gamma),
\label{eq:Hamiltonian}
\end{align}
where the first term is the translational energy ($M$: molecular mass, $\vec{R}_{i}$: center-of-mass coordinate of the guest in cage $i$) and the second is the rotational term,
\begin{align}
\hat{T}_{\mathrm{rot}}^{i} =\begin{cases}
\hfil B_{0}\hat{J}_{i}^2 &,\text{linear rotor}\\
A_{e}\hat{J}^{2}_{a,i}+B_{e}\hat{J}^{2}_{b,i}+C_{e}\hat{J}^{2}_{c,i} &,\text{asymmetric top}~,
\end{cases} 
\label{eq:Trot}
\end{align}
with rotational constants $B_{0}$ (linear) or $A_{e},B_{e},C_{e}$ (asymmetric top) and angular momentum operators $\hat{\mathbf{J}}_{i}$. Translations $\vec{R}$ and rotations $\Omega$ are coupled by the guest-cage potential $\hat{V}_{\mathrm{guest-cage}}^{i}$, for which we use analytic pairwise-additive forms~\cite{serwatka2023endo}. The guests interact via dipole-dipole couplings,
\begin{align}
\hat{V}_{\mathrm{dd}}^{ij}=\mu^2_{\mathrm{eff}}\left(\frac{\hat{e}_{i}\hat{e}_{j}-3(\hat{r}_{ij}\hat{e}_{i})(\hat{r}_{ij}\hat{e}_{j})}{L_{ij}^3}\right),
\end{align}
where $\hat{e}_{i}(\Omega_i)$ is the dipole orientation in cage $i$, $\hat{r}_{ij}$ the unit vector connecting guests $i$ and $j$, and $L_{ij}=|\vec{R}_i-\vec{R}_j|$ their separation. Parametrically the potential depends on the cage spacing $R_0$ and angle $\gamma$ (Fig.~\ref{fig:zigzag}); $L_{ij}\neq R_0$ in general since guests translate within their cages. The interaction strength is set by the effective dipole $\mu_{\mathrm{eff}}=\mu_{0}/\sqrt{\varepsilon_{r}^{\mathrm{cage}}}$, with $\mu_0$ the free-molecule dipole and $\varepsilon_{r}^{\mathrm{cage}}$ the cage screening. We restrict the second sum in Eq.~\eqref{eq:Hamiltonian} to nearest (NN) and next-nearest neighbors (NNN).

We expand the ground state of Eq.~\eqref{eq:Hamiltonian} in a tensor product basis,
\begin{align}
\ket{\Psi}=\sum_{\sigma_{1},...,\sigma_{N}}C_{\sigma_{1}\cdots \sigma_{N}}\Ket{\sigma_{1}}\otimes\cdots\otimes\Ket{\sigma_{N}}.
\label{eq:wf}
\end{align}
The local basis $\lbrace\Ket{\sigma_{i}}\rbrace$ consists of the monomer eigenstates of $\hat{H}_{\mathrm{mono},i}=-\frac{1}{2M}\Delta_{\vec{R}_{i}}+\hat{T}_{\mathrm{rot},i}+\hat{V}_{\mathrm{guest-cage}}^{i}$ \cite{serwatka2023endo}, and the expansion tensor is written as a matrix product state (MPS),
\begin{align}
C_{m_{1}\cdots m_{N}}&=\sum_{\alpha_{1},...,\alpha_{N-1}}A_{1\alpha_{1}}^{m_{1}}\cdots A_{\alpha_{N-1}1}^{m_{N}}\notag\\
&= \mathbf{A}^{m_{1}}\cdots\mathbf{A}^{m_{N}},
\label{eq:MPS}
\end{align} 
with open boundary conditions. The dimensions of the matrices $\mathbf{A}^{m}$ are the bond dimensions, which can typically be kept far below the value required algebraically by Eq.~\eqref{eq:MPS}, yielding a compact yet accurate representation \cite{schollwock2011density}. We optimize the MPS with DMRG~\cite{white1992density}. A central quantity is the von-Neumann entanglement entropy,
\begin{align}
S_{\mathrm{vN}} = -\mathrm{tr}\left(\hat{\rho}_{A}\ln\left(\hat{\rho}_{A}\right)\right)
\end{align}
where $\hat{\rho}_{A}$ is the reduced density operator of subregion $A$, measuring its entanglement with the remainder $B$. In one-dimensional gapped systems $S_{\mathrm{vN}}$ obeys an area law \cite{hastings2007area,eisert2010colloquium} whose violation near criticality diagnoses quantum phase transitions \cite{calabrese2004entanglement,pollmann2010entanglement}. During the DMRG sweeps the cut between $A$ and $B$ is moved across the chain, with iterative diagonalization and singular value decomposition selecting bond dimensions large enough to capture this entanglement faithfully (see Refs.~\onlinecite{schollwock2011density,serwatka2022ground}).

We consider the linear rotor LiF and the asymmetric top o-D$_2$O, both of which form ferroelectric phases in linear chains~\cite{serwatka2023endo}. For LiF we use the screened dipole moment $\mu_{\mathrm{eff}}=\SI{1.46}{D}$ from electronic structure calculations~\cite{serwatka2023endo}, which supports order. The screened o-D$_2$O value ($\mu_{0}=\SI{0.51}{D}$~\cite{meier2015electrical}) is too small given its large rotational constants, so we use the unscreened $\mu_{0}=\SI{1.86}{D}$~\cite{shostak1991dipole}; experimentally this reduced screening could be reached by cage functionalization that removes aromaticity while preserving the weak dispersive guest-cage coupling. The cage spacing is set to \SI{10}{\angstrom}, consistent with experimental van der Waals gaps~\cite{biskupek2020bond}. We use $33$ monomer states ($j_{\mathrm{max}}=n_{\mathrm{max}}=9$) for LiF and $33$ ($29$) states for p-H$_2$O (o-D$_2$O) with $j_{\mathrm{max}}=n_{\mathrm{max}}=6$. DMRG is performed with ITensor~\cite{fishman2022itensor} at SVD and relative-energy cutoffs of $10^{-5}$. Except in the Binder-parameter runs, a small site field $-\vec{F}\cdot(\mu_{\mathrm{eff}}\hat{e})=-\SI{0.1}{\per\centi\metre}$ with $\vec{F}=(0,F_y,F_z)^{\intercal}$ breaks inversion and reflection symmetry.

\begin{figure}
\includegraphics[width=\columnwidth]{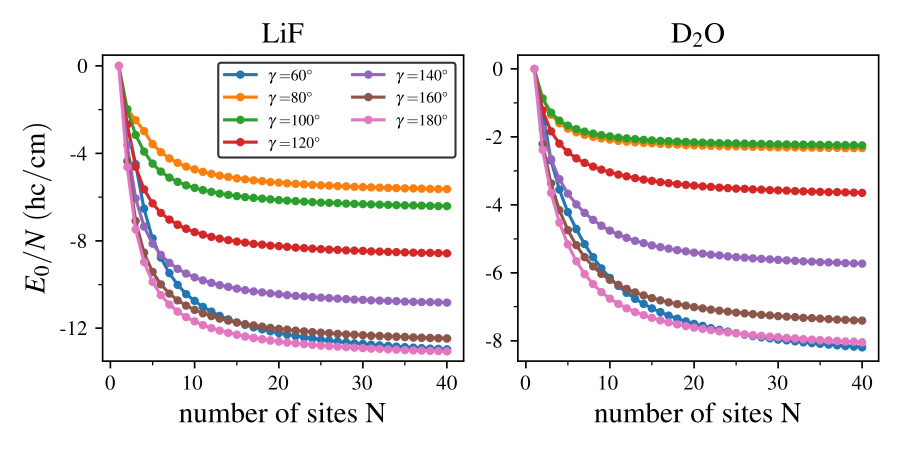}
\includegraphics[width=\columnwidth]{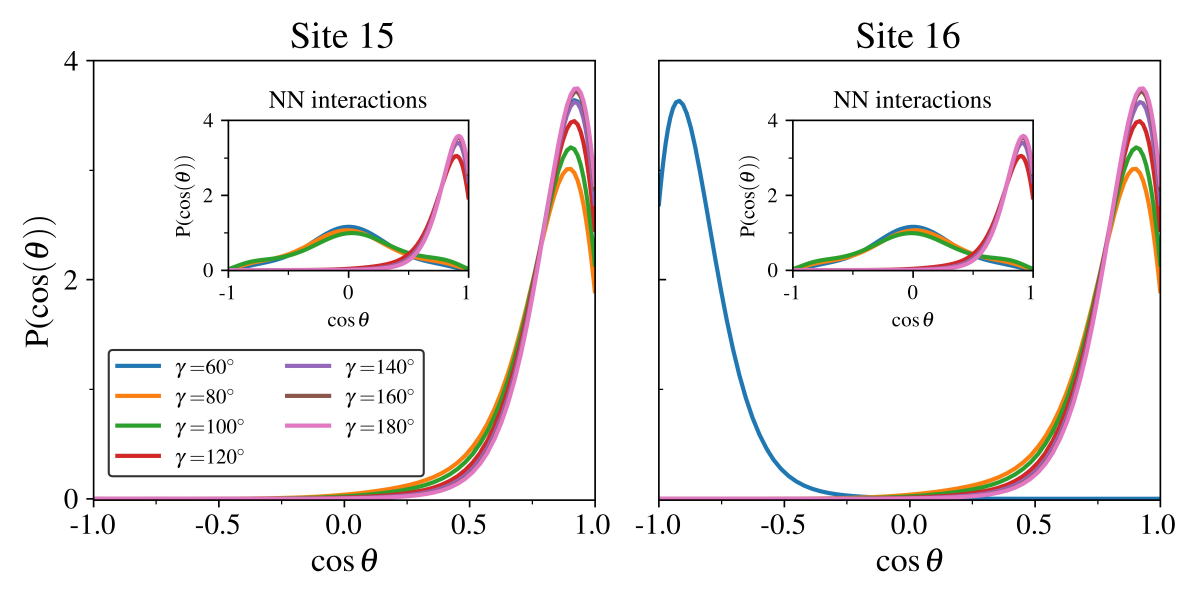}
\caption{(Upper panel) Energy per molecule for LiF@C$_{60}$ chains with different $\gamma$. (Lower panel) Density distribution along $\cos(\theta)$ of the two central sites in LiF@C$_{60}$ chains with $N=30$ and different $\gamma$. Insets: density distribution calculated for similar chains with only NN interactions.}
\label{fig:energy_density}
\end{figure}

We first probe chain stability via the energy per cage $E_0/N$ (zero referenced to separated cages), shown in Fig.~\ref{fig:energy_density}. LiF@C$_{60}$ and o-D$_2$O@C$_{60}$ behave similarly: $E_0/N$ nearly saturates by $N=20$--$30$ and, at fixed $N$, rises (less stable) with increasing $\gamma$. This trend reverses at $\gamma=60^\circ$, where the chain is as stable as the linear one, while intermediate angles $\gamma \notin \{60^\circ,180^\circ\}$ require longer chains for $E_0/N$ to saturate. To explain this restabilization we examine the single-site planar orientational density $P(\cos\theta)$ of the two central sites of $N=30$ chains (Fig.~\ref{fig:energy_density}, lower panel).

The polar angle $\theta$ is measured between the chain axis $z$ and the dipole (Fig.~\ref{fig:zigzag}). For $\gamma>80^\circ$ neighboring molecules both localize near $\cos\theta=1$, giving the ferroelectric phase known from linear chains~\cite{serwatka2022ferroelectric,serwatka2023quantum,serwatka2023endo}; the distributions broaden with increasing $\gamma$ as parallel alignment becomes less favorable. At $\gamma=60^\circ$ the next site instead localizes near $\cos\theta=-1$: neighbors anti-align and the system develops antiferroelectric N\'{e}el order, accounting for the energy drop in Fig.~\ref{fig:energy_density}. Capturing this reorientation requires extending the usual NN approximation to include NNN interactions, since at $\gamma=60^\circ$ the three NN form an equilateral triangle so that the NNN distance, which controls the dipolar coupling strength, equals the NN one. If only NN interactions are kept, the dipoles instead form antialigned pairs perpendicular to the chain ($\cos\theta\approx0$) for $\gamma\leq100^\circ$, as seen in the single-site densities in the insets of Fig.~\ref{fig:energy_density}, and the antialignment sets in at much larger $\gamma$ than with NNN included. This can be understood from the fact that $V_{ij}$ is minimized when sites $i$ and $j$ align along the connecting vector $\hat{r}_{ij}$. When NNN are included, the NNN bonds can always be satisfied by alignment along the chain axis, so for $\gamma\in[60^\circ,100^\circ]$ it is energetically favorable to satisfy the NNN bonds at the cost of frustrating the NN ones: minimizing the NN interactions at these angles would instead frustrate both NN and NNN and raise the total energy. At $\gamma=60^\circ$ the NNN still prefer alignment, but it now becomes favorable to anti-align the NN, driving the rotors to locally reorient as $\gamma$ decreases, provided the interactions are strong enough for order to develop. 

\begin{figure}
\includegraphics[width=\columnwidth]{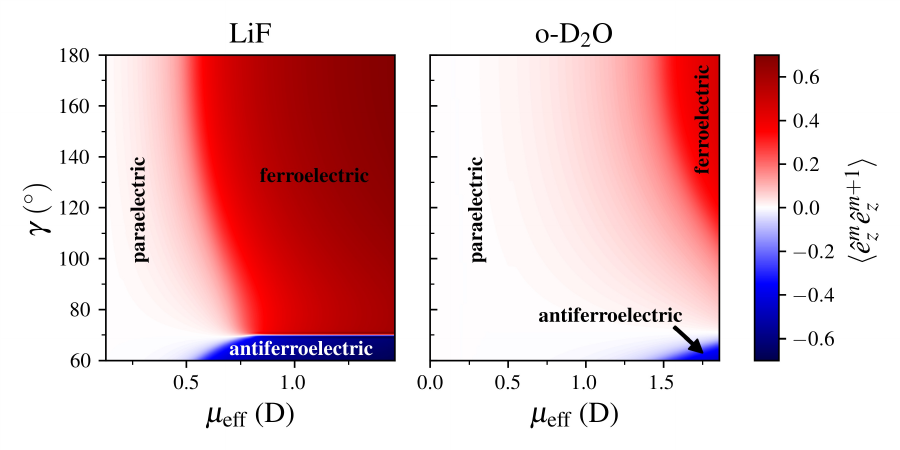}
\caption{Phase diagram of a LiF@C$_{60}$ chain and a D$_2$O chain with $N=20$ each.}
\label{fig:phases}
\end{figure}

This reorientation marks a quantum phase transition distinct from the paraelectric-ferroelectric (order-disorder) transition studied previously~\cite{serwatka2023endo}: here, at sufficiently large dipole moment, the chain angle $\gamma$ drives a transition between two ordered phases (ferroelectric and antiferroelectric). To exhibit this we plot the NN correlation function versus $\mu_{\mathrm{eff}}$ and $\gamma$ in Fig.~\ref{fig:phases}.
The NN correlation changes sign between ferroelectric and antiferroelectric order, making it a good phase marker. Fig.~\ref{fig:phases} shows that the critical dipole moment grows as $\gamma$ moves away from the optimal alignment angles---$180^\circ$ (parallel) and $60^\circ$ (antiparallel)---in either regime. The ferroelectric region is markedly larger than the antiferroelectric one, plausibly because the NNN distance $\propto \sin(\gamma/2)$ varies faster near $\gamma=60^\circ$ than near $180^\circ$.

\begin{figure}
\includegraphics[width=\columnwidth]{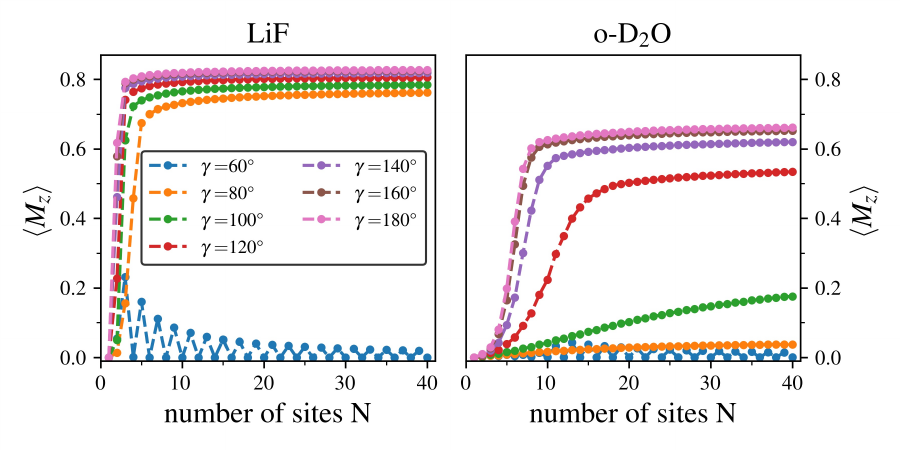}
\includegraphics[width=\columnwidth]{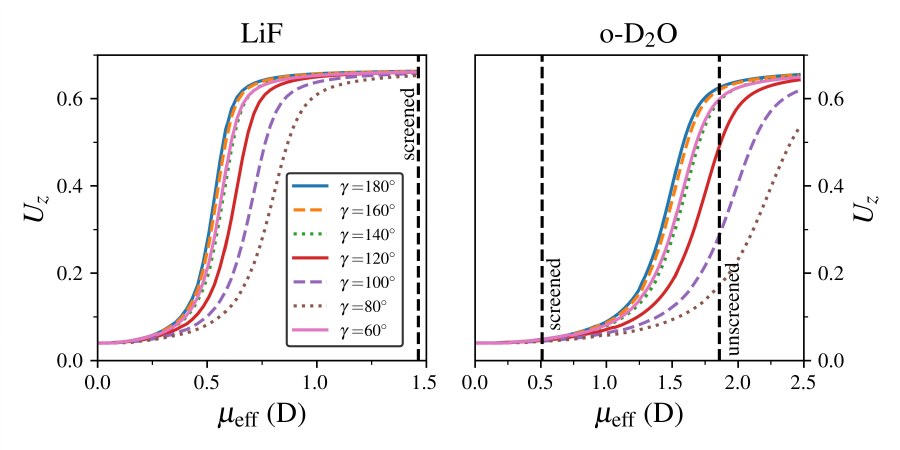}
\includegraphics[width=\columnwidth]{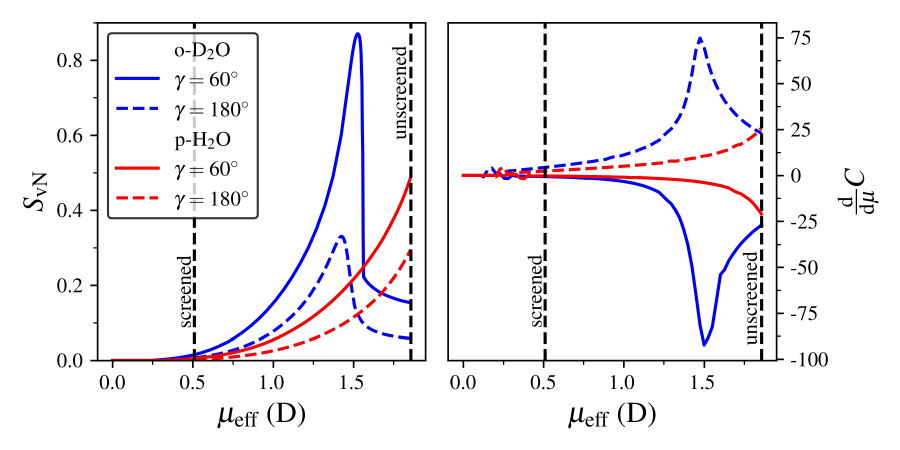}
\caption{(Upper panel) Axial, scaled polarization per cage of LiF@C$_{60}$ and o-D$_{2}$O chains at $\mu_{\mathrm{eff}}=\mu_{\mathrm{unscreened}}$. (Middle panel) Binder parameter of a LiF@C$_{60}$ chain and a o-D$_2$O chain with $N=10$ each. The Binder parameter is defined as $U_{z}=1-\frac{\langle M^4_{z}\rangle}{3\langle M_{z}^2\rangle^2}$. As operator we take the polarization operator $M_{z}=\sum_{i=1}^{N}z_{i}$ for all $\gamma>60^\circ$ and the staggered polarization operator $M_{z}=\sum_{i=1}^{N}(-1)^{i}z_{i}$ for $\gamma=60^\circ$. (Lower panel) von-Neumann entanglement entropy (left) and derivative of the correlation function (right) for linear and bent p-H$_2$O and o-D$_2$O chains.}
\label{fig:polarization_binder}
\end{figure}

The two ordered phases differ in their net polarization, shown versus chain length in the top panel of Fig.~\ref{fig:polarization_binder}.
We use the scaled axial polarization per cage, so unity marks perfectly aligned dipoles. All LiF chains with $N>5$ and $\gamma\geq80^\circ$ reach $0.7$--$0.8$, decreasing with $\gamma$ as alignment worsens. At $\gamma=60^\circ$ the curve oscillates with $N$ due to the even/odd dipole imbalance, which decays as $N\to\infty$. o-D$_2$O follows the same trends but with a faster drop in polarization with decreasing $\gamma$, slower convergence (e.g.\ at $\gamma=100^\circ$), and smaller magnitude overall---near zero at $\gamma=80^\circ$ for all $\mu_{\mathrm{eff}}$ simulated. These differences are clarified by the Binder parameter (middle panel of Fig.~\ref{fig:polarization_binder}).

The Binder cumulant vanishes in the disordered phase and saturates in the ordered one, with its inflection point locating the transition. For LiF, all $\gamma$ lie deep in the ordered phase even at the screened dipole; the transition shifts to larger $\mu_{\mathrm{eff}}$ as $\gamma$ decreases but stays well below the screened LiF value. For o-D$_2$O the same shift occurs, but since the transition sits close to the unscreened dipole, it eventually crosses it, leaving the system disordered for $60^\circ<\gamma\leq100^\circ$ (essentially no order at $\gamma=80^\circ$). This explains the small polarization and slow convergence above, and matches the weak NN correlation seen in Fig.~\ref{fig:phases} for $70^\circ<\gamma<100^\circ$. 
We have also confirmed that the Binder cumulant shows first-order behavior when crossing the transition between the antiferroelectric and ferroelectric phases in Fig.~\ref{fig:phases}.

Previously, linear p-H$_2$O@C$_{60}$ chains showed no ordered phase even with an unscreened dipole, requiring deuteration to reduce the rotational constants~\cite{serwatka2023endo}. Since bent chains at $\gamma=60^\circ$ give most sites four NN rather than two, the added equal interactions might enable ordering in p-H$_2$O. We test this with $N=30$ ground states at $\gamma=180^\circ$ and $60^\circ$, comparing the von-Neumann entropy \cite{entanglement_kitaev, entanglement_review} of p-H$_2$O and o-D$_2$O (lower panel of Fig.~\ref{fig:polarization_binder}).
o-D$_2$O shows an entropy peak---marking the transition---for both linear and bent chains, confirming ordered phases. p-H$_2$O shows no area-law violation in the relevant $\mu_{\mathrm{eff}}$ range, hence no order in either geometry. The derivative of the NN correlation (lower-right panel) peaks for o-D$_2$O with opposite signs at $\gamma=180^\circ$ and $60^\circ$, reflecting ferro- and antiferroelectric order; p-H$_2$O shows the same sign pattern, indicating prealignment, but no peaks since its dipole is too small. The only effect of the extra NN in the bent chain is a higher entropy, as expected from the area law, since the half-chain boundary is crossed by two bonds at $\gamma=60^\circ$ versus one at $\gamma=180^\circ$.

Using DMRG, we studied LiF and o-D$_2$O endofullerene chains over a range of bend angles $\gamma$. For large enough $\mu_{\mathrm{eff}}$ we find ferroelectric phases for $60^\circ<\gamma<180^\circ$, with the transition shifting to larger $\mu_{\mathrm{eff}}$ as $\gamma$ decreases. Since linear o-D$_2$O is near criticality~\cite{serwatka2023endo}, bending drives it from ferroelectric to paraelectric, whereas LiF stays ordered for $\gamma>60^\circ$. Ground-state stability decreases with $\gamma$ down to $60^\circ$, where the equilateral chain restabilizes into an antiferroelectric N\'{e}el phase with dipoles anti-aligned along the axis---showing that cage arrangement controls guest alignment. A bent chain is the narrowest stripe of a two-dimensional material; future work on true 2D endofullerene lattices, fabricable via molecular surgery and layer synthesis~\cite{cho2012structural,hou2022synthesis,meirzadeh2023few,murata2008surgery,bloodworth2022synthesis}, should host a far richer variety of quantum phases.

Comparing p-H$_2$O and o-D$_2$O, the entropy rises from linear to equilateral chains in both, yet p-H$_2$O remains disordered: even the equilateral stripe cannot overcome its large rotational constants. A 2D or 3D lattice may be needed to order water, as hypothesized for fullerene crystals~\cite{cioslowski1992endohedral,aoyagi2013cubic}. This parallels the long-debated question of ferroelectric ordering in confined and low-dimensional water and ice \cite{bramwell1999ferroelectric,iedema1998ferroelectricity,luo2008ferroelectric,zhao2014ferroelectric,koga2001formation,wang2021abnormal}, where confinement geometry, hydrogen bonding, and large rotational constants continue to challenge experiment and theory.
\begin{acknowledgments}
\section*{acknowledgments}
The authors acknowledge the Natural Sciences and Engineering Research Council (NSERC) of Canada (RGPIN-03725-2022), the Ontario Ministry of Research and Innovation (MRI), the Canada Research Chair program (950-231024), the Digital Research Alliance of Canada, and the Canada Foundation for Innovation (CFI) (project No. 35232). T. S. acknowledges a Walter-Benjamin funding of the Deutsche Forschungsgemeinschaft (Projektnummer 503971734). 
Research at the Perimeter Institute is supported in part by the Government of Canada through the Department of Innovation, Science and Economic Development Canada and by the Province of Ontario through the Ministry of Economic Development, Job Creation and Trade.
\end{acknowledgments}

\begin{acknowledgments}
\emph{Data availability.}---The data and DMRG code that support the findings of this study are openly available in the Zenodo repository at https://doi.org/10.5281/zenodo.21271590.
\end{acknowledgments}

\end{document}